\documentclass[twocolumn,amsmath,showpacs,amsfonts,aps,prc,floatfix]{revtex4}
\usepackage{graphicx}
\usepackage{bm}

\begin{document}

\title{Transverse momentum spectra and elliptic flow    in ideal hydrodynamics   and geometric scaling }
\author{Victor Roy}
\email[E-mail:]{victor@veccal.ernet.in}
\affiliation{Variable Energy Cyclotron Centre, 1/AF, Bidhan Nagar, 
Kolkata 700~064, India}
\author{A. K. Chaudhuri}
\email[E-mail:]{akc@veccal.ernet.in}
\affiliation{Variable Energy Cyclotron Centre, 1/AF, Bidhan Nagar, 
Kolkata 700~064, India}

\begin{abstract}
In an ideal hydrodynamic model, with an equation of state where the confinement-deconfinement transition is a cross-over at $T_{co}=196 MeV$,  we have simulated $\sqrt{s}$=200 GeV Au+Au collisions. 
Simultaneous description of the experimental charged particle's $p_T$ spectra and elliptic flow require that in central (0-10\%) Au+Au collisions, initial energy density scales with the binary collision number density. In less central collisions, experimental data demand scaling with the participant density. Simulation studies also indicate that in central  collisions  viscous effects are minimal.
\end{abstract}

\pacs{12.38.Mh  ,47.75.+f,  25.75.Ld} 

\date{\today}  

\maketitle

 


One of the important findings in Au+Au collisions at RHIC is that the centrality dependence of particle multiplicity can be understood in a simple geometric model. For example in \cite{Kharzeev:2000ph},    it was assumed that a fixed fraction $x$ of multiplicity per unit (pseudo) rapidity $n_{pp}=\frac{dN_{ch}}{d\eta}$ in 
pp collisions is due to hard 'processes' and the rest $(1-x)$ is due to soft processes.  Assuming that hard processes scales with the binary collision numbers
($N_{coll}$) and the soft processes scales with the participant numbers ($N_{part}$), the pseudo-rapidity density in nucleus-nucleus collisions is then parameterised as, 
 
\begin{equation} \label{eq1}
\frac{dN}{d\eta} = n_{pp}[(1-x) \frac{N_{part}}{2}+ x N_{coll}]
\end{equation}

   Using Glauber model to compute $N_{part}$ and $N_{coll}$, PHOBOS data  \cite{Back:2000gw} on charged particle multiplicity, in   Au+Au collisions were fitted with hard scattering fraction $x=0.05\pm 0.03$ at $\sqrt{s}$=56 GeV and  $x=0.09 \pm 0.03$ at $\sqrt{s}$=130 GeV \cite{Kharzeev:2000ph}. More recently,  PHOBOS collaboration studied the geometric scaling of (pseudo) rapidity density in $\sqrt{s}$=19.6 and 200 GeV Au+Au collisions \cite{Back:2004dy}. For both the collision energies, hard scattering fraction is approximately constant, $x=0.13 \pm 0.01(stat) \pm 0.05 (sys)$. 
 
 Geometric scaling of Au+Au collisions as in Eq.\ref{eq1}, has been widely used in hydrodynamic model calculations \cite{Kolb:2001qz,QGP3,Hirano:2009ah}. Hydrodynamic models require initial energy density configuration.  
Following Eq.\ref{eq1}, in an impact parameter {\bf b} Au+Au collision, initial energy density   in the transverse plane  can be parameterised as,

\begin{equation}\label{eq2}
\varepsilon({\bf b},x,y)=\varepsilon_0 [(1-x) N_{part}({\bf b},x,y)+xN_{coll}({\bf b},x,y)],
\end{equation}  

\noindent where $N_{part}({\bf b},x,y)$ and $N_{coll}({\bf b},x,y)$ are the transverse density distribution for the participant pairs and the collision number respectively. $x$ is the fraction of hard scattering. Parameterisation Eq.\ref{eq2} is generally called Glauber model initialisation. $\varepsilon_0$ in Eq.\ref{eq2} is the central energy density in ${\bf b}=0$ collision. It is generally fitted to experimental data, e.g. multiplicity, $p_T$ spectra etc. 
  Ideal hydrodynamics, with Glauber model initialisation, with hard scattering fraction $x=0.25$ explains a variety of experimental data, e.g.  identified particle's multiplicity, mean momentum, $p_T$ spectra, elliptic flow etc \cite{QGP3}. 
Ideal hydrodynamics description to the experimental data however deteriorates beyond $p_T \approx$1.5 GeV,  presumably due to increasing role of dissipative effects at large $p_T$. Description also deteriorates in very peripheral collisions. Glauber model initialisation of the energy density, with $x=0.13$, also   give reasonable description to the   
experimental data \cite{Hirano:2009ah}. One inconsistency however remained. 
Glauber model initialisation with hard scattering fraction $x=0.25$ or $x=0.13$,  
 {\em under predict} experimental elliptic flow in   very central, e.g. 0-10\% collisions.

Elliptic flow is a key observable in establishing that in $\sqrt{s}$=200 GeV Au+Au collisions at RHIC, lattice QCD predicted  Quark-Gluon-Plasma (QGP) is produced. QGP is a thermal system. Finite elliptic flow in Au+Au collisions and the fact that hydrodynamic models do explain the flow are generally cited as proof of QGP production in   Au+Au collisions.   It is then  important to understand why Glauber model of initial condition underestimate elliptic flow   in very central collisions. Neglect of dissipative effects can not be the reason. Inclusion of dissipative effects only reduces elliptic flow.

To understand the relation between elliptic flow in central Au+Au collisions and
the geometric scaling of initial energy density as in Eq.\ref{eq2}, we have simulated $\sqrt{s}$=200 GeV Au+Au collisions with
Glauber model initial condition at two extreme limits of the  hard scattering fraction, $x=0$ and $x=1$. Two limits corresponds to very different collision dynamics, for $x=0$,   initial energy density   scales with participant density implying that Au+Au collisions are completely dominated by the soft processes.
while for $x=1$, Au+Au collisions are completely dominated by the hard processes and the initial energy density scales with the density of binary collision numbers.  Actual scenario may be in between the two extreme limits. 

Details of a hydrodynamic model can be found in \cite{QGP3}. 
We assume that in Au+Au collisions, a 'baryonless', 'ideal' QGP fluid is produced.
Space-time evolution of the fluid is obtained by solving the 
energy-momentum conservation equation, 

\begin{equation} \label{eq3}
  \partial_\mu T^{\mu\nu}=0.  
  \end{equation}
  
\noindent with the code  'AZHYDRO-KOLKATA'   \cite{Chaudhuri:2008ed,Chaudhuri:2008sj,Chaudhuri:2008je} in
  ($\tau=\sqrt{t^2-z^2}$, x,y,$\eta=\frac{1}{2}\ln\frac{t+z}{t-z}$) co-ordinate system.  Longitudinal boost-invariance is assumed. Eq.\ref{eq3} is closed with an equation of state (EOS), $p=p(\varepsilon)$. We use the recently constructed lattice+HRG EOS \cite{Chaudhuri:2009uk} where the confinement-deconfinement transition is a cross-over at $T_{co}$=196 MeV.   The deconfined part of the EOS corresponds to  recent lattice simulations \cite{Cheng:2007jq}, while the confined part   corresponds to that of a hadronic resonance gas with all the resonances with mass $m_{res} \leq $2.5 GeV.   In \cite{Chaudhuri:2009uk} it was shown that the lattice based EOS with confinement-deconfinement cross-over transition, reasonably well explain the centrality dependence of $\phi$ mesons multiplicity ($dN^\phi/dy$), mean $p_T$ ($\langle p_T^\phi\rangle$) and integrated elliptic flow ($v_2^\phi$). Recently, it was shown that hydrodynamical evolution of QGP fluid, with the lattice+HRG EOS reasonably well explains the transverse momentum spectra of $\pi$, $K$ and $\phi$ mesons \cite{Roy:2010qg}. Proton spectra however are under predicted. 
  
Solution of Eq.\ref{eq3} require initial conditions, e.g. the initial or the thermalisation time $\tau_i$ beyond which hydrodynamics is applicable,
initial energy density, fluid velocity etc.  We assume that the fluid is thermalised in the time scale $\tau_i$=0.6 fm \cite{QGP3}. At $\tau_i$=0.6 fm,   energy density in the transverse plane is distributed as in Eq.\ref{eq3}, with hard scattering fraction (i) $x=0$ or (ii) $x=1$. Irrespective of the hard scattering fraction, initial fluid velocity is assumed to be zero, $v_x(x,y)=v_y(x,y)=0$. Hydrodynamic models also require a freeze-out prescription
to convert the information about the fluid energy density and  velocity to particle spectra.  We assume that  the fluid undergoes kinetic freeze-out at temperature $T_F$=150 MeV (for baryonless fluid chemical potential is zero throughout the evolution). 

\begin{figure}[t]
\center
 \resizebox{0.35\textwidth}{!}{%
  \includegraphics{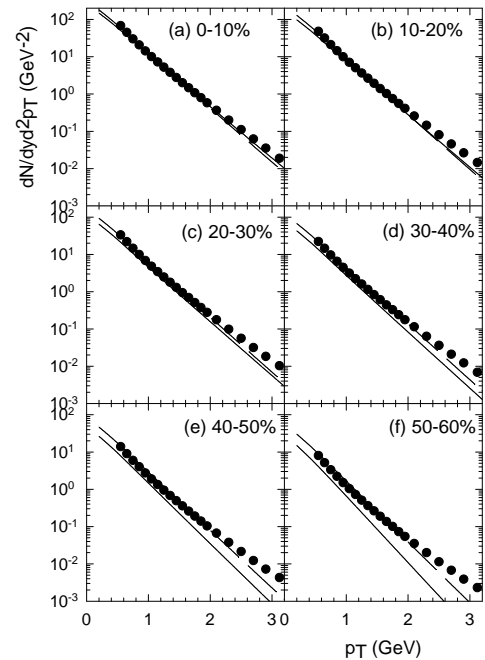}
}
\caption{In panels (a-f) PHENIX data \cite{Adler:2003au} on the charged particle transverse momentum spectra in 0-60\% Au+Au collisions are compared with hydrodynamic model predictions. The dashed lines are the predictions when initial energy density scales with participant density ($x=0$). The solid line are predictions when energy density scales with density distribution of binary collision numbers ($x=1$).  }
  \label{F1}
\end{figure}

With hard scattering fraction fixed, either $x=0$ or $x=1$, central energy density $\varepsilon_0$ is the only parameter left in the model. We fix $\varepsilon_0$ by  fitting the  PHENIX data on
charged particles $p_T$ spectra in 0-10\% Au+Au collisions \cite{Adler:2003au}. 
For $x=0$, best fit to the 0-10\% data is obtained with  $\varepsilon_0$=36.1 $GeV/fm^3$. 
PHENIX data require $\sim$30\% higher central energy density,
$\varepsilon_0$=48 $GeV/fm^3$, if 
 Au+Au collision is completely dominated by the hard processes ($x=1$).   
 The solid and the dashed lines in Fig.\ref{F1}a  are fit to the data with hard scattering fraction $x=1$ and $x=0$ respectively. Data are well fitted. 
Charged particles  $p_T$ spectra in central  Au+Au collision is insensitive to the hard scattering fraction $x$ in the Glauber model of initial condition.

With initial energy density fixed, we can predict for the $p_T$ spectra   in all the other collision centralities. 
In Fig.\ref{F1}, in  panels (b-f),   model  predictions for the charged particles $p_T$ spectra in 10-20\%, 20-30\%, 30-40\%, 40-50\% and 50-60\% Au+Au collisions are compared with the PHENIX data \cite{Adler:2003au}.  
When central energy density is fixed to reproduce charged particles $p_T$ spectra in 0-10\% collisions, Glauber model initialisation with $x=1$,  also give reasonable description to the spectra in 10-20\% and 20-30\% centrality collisions.  But in more peripheral collisions, PHENIX data are under predicted.
Initialisation with hard scattering fraction $x=0$ however continue to explain data till very peripheral collisions, though description to the data deteriorates at larger $p_T$ or in more peripheral collisions.
It appear that if in Au+Au collisions, energy density scales with participant density ($x=0$), charged particles $p_T$ spectra in 0-60\% Au+Au collisions are reasonably well explained. On the contrary, if energy density scales with binary collision number density ($x=1$), charged particles $p_T$ spectra, only in   central (0-30\%) collisions is explained. 

\begin{figure}[t]
\center
 \resizebox{0.35\textwidth}{!}{%
  \includegraphics{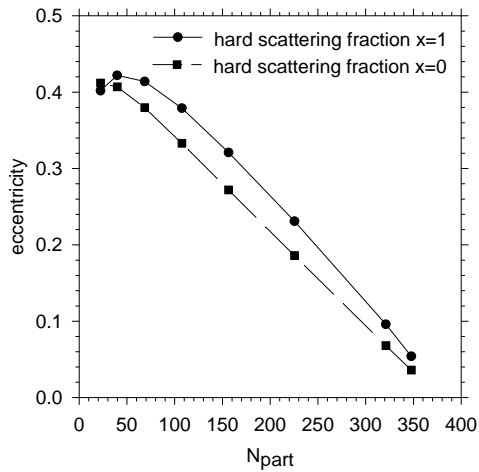}
}
\caption{Initial eccentricity of the collision zone as a function of participant numbers. The filled circles and squares are eccentricity with hard scattering fraction 0 and 1 respectively.}
  \label{F2}
\end{figure}

Let us now study centrality dependence of simulated elliptic flow.
  In a hydrodynamic model, elliptic flow depends on the initial spatial eccentricity, 
$\varepsilon_x=\frac{\langle y^2-x^2\rangle}{\langle x^2+y^2\rangle}$,
${\langle ...\rangle}$ denotes energy density weighted averaging. 
 In Fig.\ref{F2}, centrality dependence of   $\varepsilon_x$, in the two extreme
  limits  $x=0$ (filled circles) and $x=1$ (filled squares) are shown.  $\varepsilon_x$   is more if Au+Au collisions are dominated by the hard processes ($x=1$) rather than the soft processes ($x=0$). Glauber model initialisation of energy density with hard scattering fraction $x=1$ will generate more elliptic flow than the initialisation with $x=0$.

\begin{figure}[t]
\center
 \resizebox{0.35\textwidth}{!}{%
  \includegraphics{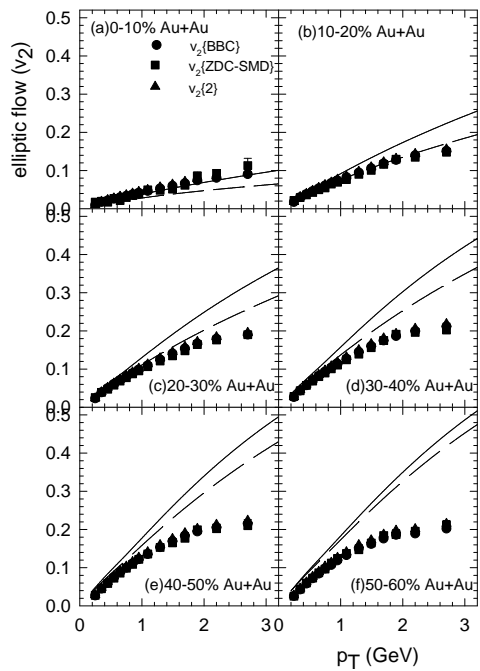}
}
\caption{ The filled circles, squares and triangles are PHENIX measurements \cite{Afanasiev:2009wq} for elliptic flow in 0-60\% Au+Au collisions. The solid and dashed  lines are elliptic flow in hydrodynamic simulations with hard scattering fraction x=1 and x=0 respectively.}  \label{F3}
\end{figure}

In Fig.\ref{F3}, we have compared the   simulated $v_2$ with PHENIX measurements \cite{Afanasiev:2009wq}. PHENIX collaboration obtained $v_2$ from two independent analysis,
(i) event plane method from two independent subdetectors, $v_2\{BBC\}$ and 
$v_2 \{ZDC-SMD\}$ and (ii) two particle cumulant $v_2\{2\}$.  $v_2\{2\}$ from two particle cumulant and $v_2\{BBC\}$ or $v_2\{ZDC-BBC\}$ from event plane methods agree within the systematic error. All the measurements upto $p_T=3$ GeV are shown in Fig.\ref{F3}.  The solid and dashed lines in Fig.\ref{F3} corresponds to Glauber model initial conditions with $x=1$ and $x=0$  respectively. As expected from the eccentricity study (Fig.\ref{F2}), in all the collision centralities,  Glauber model initialisation with $x=1$ generate more flow that the initialisation with $x=0$.   Unlike the $p_T$ spectra in central Au+Au collisions, which do not distinguish between the initial conditions with $x=1$ and $x=0$,  elliptic flow, being a more sensitive observable, can distinguishes
between them.
It is very interesting to note that Glauber model initialisation with hard scattering fraction $x=1$  well explain the 
PHENIX data on elliptic flow in 0-10\% Au+Au collisions. However, in all the other collision centralities elliptic flow is  over predicted.  For example, at $p_T\approx$1.5 GeV, simulated flow with $x=1$ over predict experiments by $\sim$20\%,  25\%, 35\%, 45\% and 60\%  in 10-20\%, 20-30\%, 30-40\%, 40-50\% and 50-60\% Au+Au collisions. At larger $p_T$ flow is even more over predicted.   
Glauber model initialisation with hard scattering fraction $x=0$ predict less flow and elliptic flow in 0-10\% Au+Au collisions is underpredicted e.g., at $p_T\sim$1.5 GeV, it under predict the experiment   by $\sim$35\%. In all the other collision centrality agreement with data is comparatively better. In 10-20\% and 20-30\% Au+Au collisions, Glauber initialisation with $x=0$ give very good description of the data up to $p_T\sim$1.5 GeV. In more peripheral collisions, flow is over predicted. Even then agreement with data is better than that obtained with hard scattering fraction $x=1$. For example at $p_T\approx$1.5 GeV, simulated flow over predict PHENIX data by $\sim$ 15\%, 25\% and 50\% in 30-40\%, 40-50\% and 50-60\% collisions. 

Considering that finite elliptic flow is an essential observable to establish that   QGP is produced in $\sqrt{s}$=200 GeV Au+Au collisions, it is imperative that hydrodynamical models, with QGP as initial fluid, explains the flow in 0-10\% Au+Au collisions in addition to the  $p_T$ spectra. Present analysis indicate that    in 0-10\% Au+Au collisions,
simultaneous description of $p_T$ spectra and elliptic flow require  
hard scattering fraction $x=1$ in the Glauber model of initial condition.
In less central collisions however, simultaneous description of $p_T$ spectra and elliptic flow are best obtained with  hard scattering fraction $x=0$. 
The result implies that geometric scaling of Au+Au collisions changes with collision centrality. In a central collision, energy density scales with binary collision number density while in a less central collision, energy density scales with participant density. Arguably, transition from  binary collision number scaling to participant scaling can not be as sharp as conjectured here. More detailed analysis is required to find the width and exact location of the transition.  

Such a transition may have implications for the hydrodynamical analysis also. 
Note that for binary collision number scaling, fluid has to be initialised at higher energy density  ($\varepsilon_0$=48 $GeV/fm^3$ for $x=1$ and 
$\varepsilon_0$=36.1 $GeV/fm^3$ for $x=0$). We have assumed similar thermalisation time $\tau_i$=0.6 fm for both the scaling conditions. 
Since thermalsation time scale is expected to be inversely proportional to the density, it is   likely that the fluid in central Au+Au collisions will thermalise in a lesser time scale that the fluid in less central collisions.
The result may also have implication on the  dynamics of the {\em pre-equilibrium} stage. Hydrodynamic models assume local thermalisation. The fluid produced in Au+Au collisions  evolves through a 
pre-equilibrium stage to equilibration. At present, we have limited knowledge about the pre-equilibrium stage. Present results suggests that
in 0-10\% Au+Au collisions, pre-equilibrium stage is dominated by binary collisions, but in a less central collision, pre-equilibrium stage is dominated by the 'wounded' nucleons. 

In the present analysis, the effect of viscosity is neglected. QGP viscosity is a contentious issue. String theory based models (ADS/CFT), which are unrelated to QCD, give a lower bound on viscosity of any matter, viscosity to entropy ratio, $\eta/s \geq 1/4\pi$ \cite{Policastro:2001yc}. Perturbative estimate of the viscosity is also uncertain to a great extent, $\eta/s\approx$ 0.0-1.0 \cite{Arnold:2000dr,Meyer:2007ic,Nakamura:2005yf}. Several authors have extracted viscosity directly from the experimental data  
\cite{Chaudhuri:2009uk,Gavin:2006xd,Drescher:2007cd,Lacey:2006bc,Chaudhuri:2009ud,Chaudhuri:2009hj,Adare:2006nq,Luzum:2008cw,Song:2008hj}. Depending on the model, data analysed etc. the extracted viscosity vary over a large range and one can possibly give an upper bound, $\eta/s < 0.5$ \cite{Luzum:2008cw,Song:2008hj}. Effect of viscosity is to enhance particle production mainly at large $p_T$, and also to reduce elliptic flow. One observes from Fig.\ref{F1} and \ref{F3}, that   in 0-10\% Au+Au collisions, if initial energy density scales with collision number density, ideal fluid dynamics hardly leave any scope for
viscous enhancement of  $p_T$ spectra or for viscous suppression of elliptic flow.
Assumption of viscous fluid evolution can only worsen the fit to the elliptic flow. We conclude that  the PHENIX data on the charged particles $p_T$ spectra and elliptic flow in 0-10\% Au+Au collisions do not demand any viscosity.
Viscosity however can be important in peripheral collisions.  In peripheral collisions, experimental $p_T$ spectra and elliptic flow are better explained
if energy density scales with participant density.  However, at large $p_T$ elliptic flow is over predicted. With viscous suppression agreement with data will be better. 

To summarise, in an ideal hydrodynamic model, we have simulated $\sqrt{s}$=200 GeV Au+Au collisions. Initial QGP fluid, thermalised at the time scale $\tau_i$=0.6 fm, evolve under the influence of an lattice+HRG equation of state, where confinement-deconfinement transition is a cross-over at $T_{co}$=196 MeV.
We have considered two initial conditions (i) initial energy density scales with the binary collision numbers and (ii) initial energy density scales with participant numbers. For both the initial conditions, initial energy density was adjusted to fit PHENIX data on charged particles $p_T$ spectra in 0-10\% collisions. It was then demanded that hydrodynamic models simultaneously explain both the $p_T$ spectra and elliptic flow in central   Au+Au collisions.   Charged particles $p_T$ spectra and elliptic flow in 0-10\% collisions are best explained if initial energy density scales with binary collision numbers. It was also indicated that   description to the data can not be improved if viscous effects are included. Less central collisions however, prefer energy density scaling with participant density. Peripheral collisions also demand viscous effects.

\end{document}